\documentclass[final,3p, times,,twocolumn]{elsarticle}

\journal{Physics Letters A}
\biboptions{sort&compress}

\usepackage{graphicx}
\usepackage{color}
\usepackage{amsmath, amssymb, graphicx, units, xspace}

\newcommand{\Ppre}{\ensuremath{\operatornamewithlimits{P_{\rm pre}}}\xspace}
\newcommand{\Pint}{\ensuremath{\operatornamewithlimits{P_{\rm int}}}\xspace}
\newcommand{\Hup}{\ensuremath{\operatornamewithlimits{H^\uparrow}}\xspace}
\newcommand{\Hdown}{\ensuremath{\operatornamewithlimits{H^\downarrow}}\xspace}

\begin{document}
\begin{frontmatter}
\title{Evaluation of selected recurrence measures in discriminating pre-ictal and inter-ictal periods from epileptic EEG data}

\author[1]{Eulalie Joelle Ngamga}
\author[2]{Stephan Bialonski}
\author[1]{Norbert Marwan}
\author[1,3,4]{J\"urgen Kurths}
\author[5a,5b]{Christian Geier}
\author[5a,5b,5c]{Klaus Lehnertz}
\address[1]{Potsdam Institute for Climate Impact Research, Telegraphenberg A 31, 14473 Potsdam, Germany}
\address[2]{Max-Planck-Institute for the Physics of Complex Systems, N\"othnitzer Stra{\ss}e 38, 01187 Dresden, Germany}
\address[3]{Department of Physics, Humboldt University Berlin, 12489 Berlin, Germany}
\address[4]{Institute for Complex Systems and Mathematical Biology, University of Aberdeen, Aberdeen AB24 3UE, United Kingdom}
\address[5a]{Department of Epileptology, University of Bonn, Sigmund-Freud-Stra{\ss}e~25, 53105~Bonn, Germany}
\address[5b]{Helmholtz Institute for Radiation and Nuclear Physics, University of Bonn, Nussallee~14--16, 53115~Bonn, Germany}
\address[5c]{Interdisciplinary Center for Complex Systems, University of Bonn, Br{\"u}hler Stra{\ss}e~7, 53175~Bonn, Germany}

\date{\today}

\begin{abstract} 
We investigate the suitability of selected measures of complexity based on recurrence quantification analysis and recurrence networks for an identification of pre-seizure states in multi-day, multi-channel, invasive electroencephalographic recordings from five epilepsy patients.
We employ several statistical techniques to avoid spurious findings due to various influencing factors and due to multiple comparisons and observe precursory structures in three patients. 
Our findings indicate a high congruence among measures in identifying seizure precursors and emphasize the current notion of seizure generation in large-scale epileptic networks.
A final judgment of the suitability for field studies, however, requires evaluation on a larger database.
\end{abstract} 

\end{frontmatter}

\section{Introduction}
Recurrence plots (RPs) are graphical representations of times during 
which two states of a system are neighbors in phase space~\cite{eckmann87}.
They have been widely used over the last 25 years as a tool to study changes and transitions in the dynamics of a system (even high-dimensional), or to detect synchronization and coupling~\cite{marwan2007recurrence,marwan2008epjst,marwan2015a}. 
This has been achieved by using the visual aspects of structures encountered in RPs as well as different statistical quantification approaches~\cite{marwan2007recurrence,marwan2008epjst}.
One important and widely used approach is {\it recurrence quantification analysis} (RQA), which is based on diagonally and vertically aligned
recurrence points in the RP~\cite{zbilut1992,marwan2002heartrate}.
These lines characterize the temporal interdependences between individual observations or segments of the phase-space trajectory. 
Several substantial measures of complexity (MOC) have been defined on the base of these temporal structures of the RP and are related to predictability, stationarity, or intermittency. 
RQA has demonstrated its potential through many successful applications in different fields~\cite{webber2015}. 
Among others, they were applied to electroencephalographic (EEG) data from epilepsy patients as well as from epileptic rats. 
For example, Acharya and colleagues~\cite{acharya2011} have used RQA measures to classify EEG data from normal, during seizures (ictal), and between seizures states (inter-ictal).
Further, RQA measures have been reported to exhibit sudden abrupt changes occurring up to some minutes before seizure onsets~\cite{thomasson2001,komala2009,zhang2008,zhu2008}. 
The latter findings can contribute to better understand this neurological disorder that affects about 65 million individuals worldwide~\cite{Thurman2011} as well as to develop alternative therapies, e.g. based on the prediction of seizures~\cite{Litt2002b,mormann2007,sackellares2008,carney2011,tetzlaff2013,lehnertz2014,Gadhoumi2015}, particularly for the 20--30~\% of patients that remain poorly treated or untreated~\cite{Schuele2008}.
It remains elusive, however, whether the described phenomena can be regarded as seizure precursors, since their statistical validity has not sufficiently or not at all been investigated, and since the analyzed EEG recordings were of rather short duration. 

Recently, another quantification approach has been introduced that combines time series recurrence with complex networks~\cite{xu2008,marwan2009,donner2010}.
Here, the recurrence matrix is considered as the adjacency matrix of an undirected and unweighted complex network. 
The resulting {\it recurrence network} (RN) can be characterized with well-known network measures, i.e., further diagnostic tools become available for time series analysis~\cite{marwan2015b,donner2015}.
In contrast to RQA, where the MOC characterize the {\it dynamical properties} of the system, these network-based measures capture the {\it geometric properties} associated with a trajectory in phase space~\cite{donner2010}. 
Such complementary information can be useful when studying regime changes~\cite{marwan2015b}, characterizing different dynamics~\cite{zou2010,gao2013}, or even for the detection of coupling directions~\cite{feldhoff2012}. 
First applications in different scientific disciplines have demonstrated the usefulness of these additional characteristics. 
Very promising findings have been discussed, e.g., by Lang and colleagues~\cite{lang2013} in a RN-analysis of synchronous EEG time series from normal subjects and from epilepsy patients. 
Among other findings, the authors observed that RNs of normal subjects exhibited a sparser connectivity and a smaller clustering coefficient compared to those of epilepsy patients (cf.~\cite{Horstmann2010}).
These findings have been confirmed by another study, reporting an increasing degree of structural complexity in the EEG of normal subjects compared to the EEG from epilepsy patients~\cite{subramaniyam2014}.

It has been suggested that the conceptual difference between RQA and RN measures may allow to capture complementary aspects of the underlying dynamics under investigation, and that the combined use of both quantification approaches may improve the detection of dynamical changes~\cite{donner2010}. 
In certain applications, however, a higher performance of RN measures in comparison to that of RQA measures has been observed. 
For example, this has been reported in Ref.~\cite{ramirez2013} where a classification of healthy women and preeclamptic patients based on cardiovascular time series has been performed with the aim of performing early prediction of preeclampsia. 
Another example is found in Ref.~\cite{zou2010}, where a classification of periodic and chaotic behavior is performed using short time series of observables from continuous-time dynamical systems. 

We observed a lack of literature describing such a comparison of performance between RQA and RN measures when applied to time series from complex systems such as the brain, and in particular in the analysis of EEG data.
In the present work, we compare selected RQA and RN measures for a specific problem of multivariate EEG data analysis. 
In particular, we investigate the suitability of these measures for an identification of  pre-seizure states in multi-day, multi-channel, invasive EEG (iEEG) recordings. 

\section{Data and methods}

\subsection{Patient characteristics and data}
\begin{figure} [htbp]
\centering\includegraphics[width=\columnwidth]{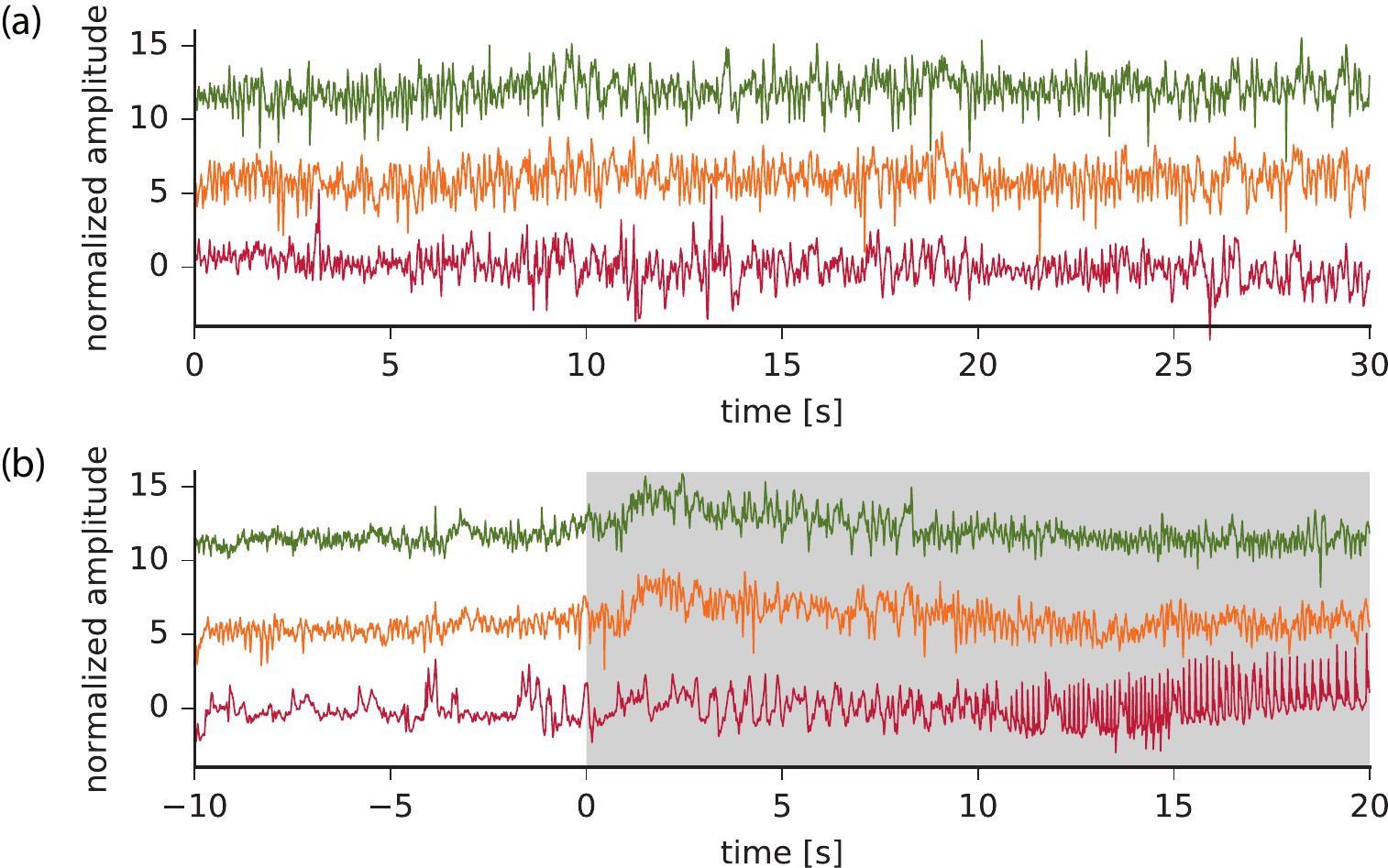}
\caption{\label{eeg_example} { Example of iEEG data for an inter-ictal (a) and a pre-ictal/ictal period (b) of patient 3 from recording sites within the epileptic focus (red), from its neighborhood (orange), and from a remote brain region (green).
The latter two time series were shifted to enhance readibility, and amplitude values were normalized to zero mean and unit variance.  The gray-shaded area marks the initial phase of the seizure.}}
\end{figure}

We analyze iEEG recordings from five epilepsy patients (see Table~\ref{tab:patdat} and {Fig.~\ref{eeg_example}}) who underwent presurgical evaluation of drug-resistant epilepsy at the University of Bonn Epilepsy Program~\cite{kral2002}. 
The patients signed informed consent that their clinical data might be used and published for research purposes. 
Further, the study protocol had received prior approval by the ethics committee of the University of Bonn. 

\begin{table}[hhh!]
 \begin{center}
 \begin{tabular}{cccccccc}
{ID} & {age/gender} & {$D_{\rm epi}$} & {FH} & {FR} & { $N_{\rm rs}$} & {$N_{\rm sz}$} & {$D_{\rm rec}$}\\ \hline
1 & 37/f &  5 & R & MT & 70 & 7 & 169\\
2 & 55/m & 10 & L & LT & 20 & 6 & 232\\
3 & 44/f & 44 & L & LT & 52 & 5 & 220\\
4 & 22/m & 19 & L & LT & 62 & 7 & 167\\
5 & 35/f &  6 & R & LT & 70 & 7 & 141\\ \hline
\end{tabular}
\end{center}
 \begin{center}\caption{Clinical data. ID: patient identification number; age (yrs.) and gender: female (f), male (m); $D_{\rm epi}$: duration of epilepsy (yrs.); FH: focal hemisphere, left (L), right (R); FR: focal region, MT mesial aspects of temporal lobe, LT lateral aspects of temporal lobe; $N_{\rm rs}$: number of recording sites; $N_{\rm sz}$: number of seizures; $D_{\rm rec}$: duration of iEEG recording (hrs.) }
\label{tab:patdat}
\end{center}
\end{table}

The iEEG data was recorded from chronically implanted intrahippocampal depth and/or subdural grid and strip electrodes (on average of 54 contacts) with a total recording time of \unit[929]{h} during which 32 seizures (five to seven seizures per patient) occurred.
The data was band-pass-filtered between 0.1 and \unit[70]{Hz}, sampled at \unit[200]{Hz} using a 16 bit analog-to-digital converter, and referenced against the average of two recording contacts outside the focal region. 
Reference contacts were chosen individually for each patient. 
Some recording gaps have been encountered and they were mainly due to diagnostic procedures
that required the patient to be temporarily disconnected from the recording system.

\subsection{RQA- and RN-based measures of complexity}

The basis of the MOC that we here used to characterize the iEEG is the recurrence plot (RP). 
It was introduced to visualize the time-dependent behavior of the dynamics of a system and particularly the recurrences of the phase-space 
trajectory to a certain state~\cite{eckmann87,marwan2007recurrence}. 
Let us consider $x$ as an exemplary univariate time series with $T$ sampling points and let $x_i$ denote the value of $x$ at discrete time $i$. 
In order to observe the recurrences of states from this time series, we compute the $T \times T$ matrix 
\begin{equation}\label{rp}
\mathbf{R}_{i,j}=\Theta(\varepsilon-|x_i - x_j|), \qquad i,j=1,\ldots,T
\end{equation}
where $\Theta(\cdot)$ is the Heaviside function, $\varepsilon$ is a predefined threshold, and $| \cdot |$ denotes absolute value. 
In general, Eq.~(\ref{rp}) can be applied on phase-space trajectories in $\mathbb{R}^m$ (where $m$ is the dimensionality and the absolute value is replaced by a norm~\cite{marwan2007recurrence}), but here we apply it on time series directly (i.e., without embedding of time series in the phase space, similarly to Ref.~\cite{goswami2013epjst}).
{ This choice is motivated by the fact that several recurrence properties are invariant under embedding~\cite{thiel2004a} 
and by the highly non-stationary character of brain dynamics~\cite{Rieke2002,Rieke2003,Rieke2004}, which complicates 
the choice of appropriate embedding parameters. Moreover,
embedding can cause spurious correlations which affect mainly the recurrence analysis of stochastic signals \cite{marwan2011}.}

\begin{figure*} [htbp]
\includegraphics[width=\textwidth]{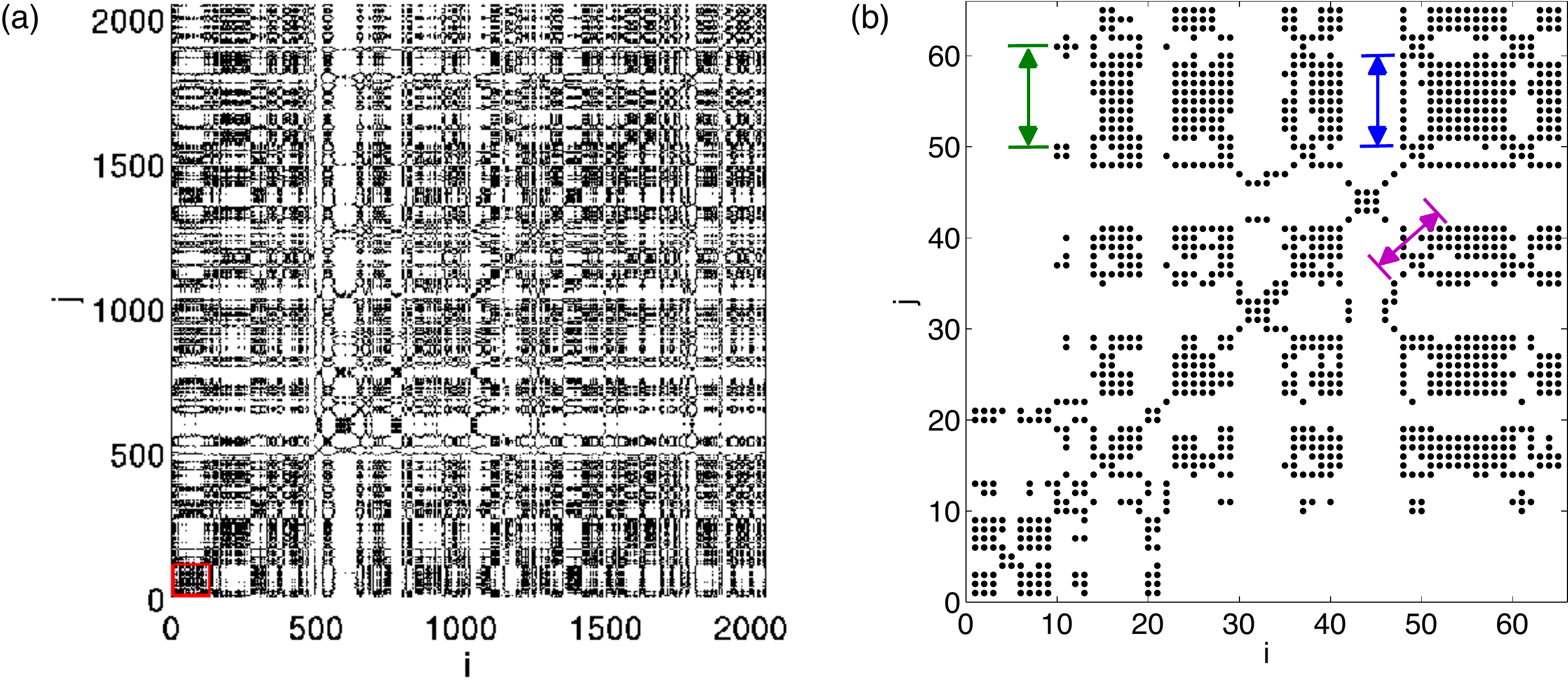}
\caption{\label{rp_illus} {(a)} Exemplary recurrence plot of \unit[10.24]{s} iEEG recording ($T = 2048$ data points). 
The corresponding recurrence matrix was calculated using the recurrence threshold $\varepsilon=0.3$. 
{(b)} Enlargement of the area marked by the red rectangle (lower left corner). 
The green, blue and magenta double arrows exemplify a white vertical, black vertical, and black diagonal line, respectively.}
\end{figure*}

An RP is a graphical representation of the above defined matrix $\mathbf{R}$. 
For the coordinate $(i,j)$ of an RP we choose black color to plot a point if $\mathbf{R}_{i,j}=1$, i.e., in the recurrent case, and white color otherwise. 
An example is shown in the left part of Fig.~\ref{rp_illus} for \unit[10.24]{s} of an iEEG recording. 
The white and black points can form different lines and structures, which are related to the properties of the underlying dynamics. 

Among these lines and structures encountered in an RP, we consider in our analysis the white vertical, black vertical, and black diagonal lines, respectively (see right part of Fig.~\ref{rp_illus}). 
These lines are commonly used in RP-based analyses~\cite{zbilut1992,marwan2007recurrence,marwan2002heartrate,webber2015}. 
The lengths of white vertical lines are an estimator of the recurrence time~\cite{ngamga2007,ngamga2012}. 
A black vertical line marks a time length in which a state does not change or changes very slowly, and is thus related to laminar states. 
A black diagonal line occurs when a segment of the trajectory runs parallel to another segment, and is thus related to the divergence of states~\cite{marwan2007recurrence}. 
By counting how many times a length of a line occurs in an RP, we evaluate the frequency distributions of the respective lengths of these three types of lines. 
From the frequency distributions, we compute the following MOC: {\it determinism} (DET), {\it laminarity} (LAM), and {\it mean recurrence time} (MRT)~\cite{marwan2007recurrence}. 

Furthermore, we use MOC that characterize complex networks, in particular the {\it average shortest path length} (APL) and {\it network transitivity} (Cl). 
The analogy between the recurrence matrix and the adjacency matrix of an undirected and unweighted complex network~\cite{marwan2009,donner2010} allows us to apply complex network measures on RPs in order to quantify the geometrical properties of the system's attractor encoded in the RP. 
For more details on this approach we refer to Refs.~\cite{marwan2009,donner2010,donges2012}.
A brief description including the mathematical equations of the MOC considered in the present work is given in the appendix. 

Using a sliding-window analysis, we calculated time profiles of the above mentioned MOC separately for each iEEG time series from each channel. 
Non-overlapping windows of 4096 data points, corresponding to a duration of \unit[20.48]{s}, were used in accordance with previous studies~\cite{mormann2005,kuhnert2010}. 
This window length can be considered as a compromise between the required statistical accuracy for the calculation of a measure and the approximate stationarity within a window's length~\cite{blanco1995,silva1987}. 
A fixed value of the threshold $\varepsilon=0.3$, Eq.~(\ref{rp}), was chosen following the discussion in Refs.~\cite{marwan2011,schinkel2008} and the data in each analysis window were normalized to zero mean and unit 
standard deviation. 

\subsection{Investigating the suitability of MOC for an identification of pre-seizure states}
\label{sec:predmeth}
We tested whether time profiles of the MOC carry potential information about seizure precursors by estimating their classification performance in terms of their ability to distinguish between inter-ictal and pre-ictal (before a seizure) periods. 
Following Mormann and colleagues~\cite{mormann2005}, we here assumed the existence of a pre-ictal period with a duration of \unit[30]{min}~\cite{elger1998,levanquyen1999}. 
In order to exclude effects from the post-ictal period (after a seizure), data recorded within \unit[1]{h} after the electrical onset of a seizure was discarded from the analysis.
Eventually, we defined inter-ictal periods to include all data except from the pre-ictal, ictal, and post-ictal periods.
In cases where the time interval between two successive seizures was less than the assumed pre-ictal duration plus one hour, the maximum amount of data available from the seizure onset back to the end of the post-ictal phase of the preceding seizure was used.
Applying these selection criteria reduced the number of seizures accessible for our analysis to 26.

For each channel, we then investigated whether the MOC amplitude values allow us to distinguish between inter-ictal and pre-ictal periods. 
To do so, we tested the separability of the pre-ictal \Ppre and inter-ictal \Pint distributions of MOC amplitude values in terms of sensitivity and specificity using receiver operating characteristic (ROC) statistics. 
With this statistics, a threshold for amplitude values is continuously shifted across \Ppre and \Pint. 
The ROC curve is obtained by plotting sensitivity (ratio of true positives to total number of positives) against one minus specificity (ratio of true negatives to total number of negatives)\footnote{ 
The definitions of sensitivity and specificity are based either on the ROC hypothesis of a {\it pre-ictal decrease} (\Hdown; MOC amplitude values from \Ppre are lower than those of \Pint) or on a {\it pre-ictal increase} (\Hup; vice versa).
For \Hdown, the terms `positive' and `negative' correspond to whether an amplitude value is below, respectively above the threshold, while the terms `true' and `false' indicate whether values below the threshold belong to \Ppre and values above the
threshold belong \Pint or not. For \Hup, these correspondences need to be adjusted accordingly.}.
The capability of the considered MOC to distinguish between inter-ictal and pre-ictal periods was then quantified using the area under the ROC curve (AUC). 
For identical distributions \Ppre and \Pint (i.e., periods are indistinguishable)
${\rm AUC} = 0.5$, while for distributions that are completely non-overlapping,
values of 0 or 1 are attained, depending on the ROC hypothesis used for the definition of sensitivity and specificity (pre-ictal decrease: ${\rm AUC} > 0.5$;
pre-ictal increase: ${\rm AUC} < 0.5$). We performed analyses for both ROC hypotheses and selected the larger one thus achieving an AUC value that is always $\geq 0.5$ by construction.

Next, in order to test whether the predictive performance -- as quantified by the largest AUC value -- is better than random, we employed the concept of seizure time 
surrogates~\cite{andrzejak2003}. 
With this Monte-Carlo-based resampling technique, the original seizure times are replaced with times randomly chosen in the inter-ictal interval (seizure time surrogates) using 
the total number of original seizures and the distribution of inter-seizure intervals as constraints. 
In addition, seizure time surrogates were not allowed to coincide with original seizure onset times or to fall into a recording gap.
For each patient and each channel, ROC analysis was repeated for 100 seizure time surrogates (the available data from patient 3 allowed to generate 20 seizure time surrogates only), and 
the significance of a given predictive performance was determined by calculating the fraction of AUC values obtained with seizure time surrogates, which exceeded AUC values obtained with the original seizure times.

Eventually, we controlled for falsely rejected null hypotheses due to multiple testings (1370 hypothesis tests for five MOC and for a total of 274 channels) by applying the Benjamini-Hochberg procedure~\cite{benjamini1995} with a false discovery rate of 0.1.

\section{Results and Discussion}

\begin{figure} [hhh!]
\centering
\includegraphics[width=\columnwidth]{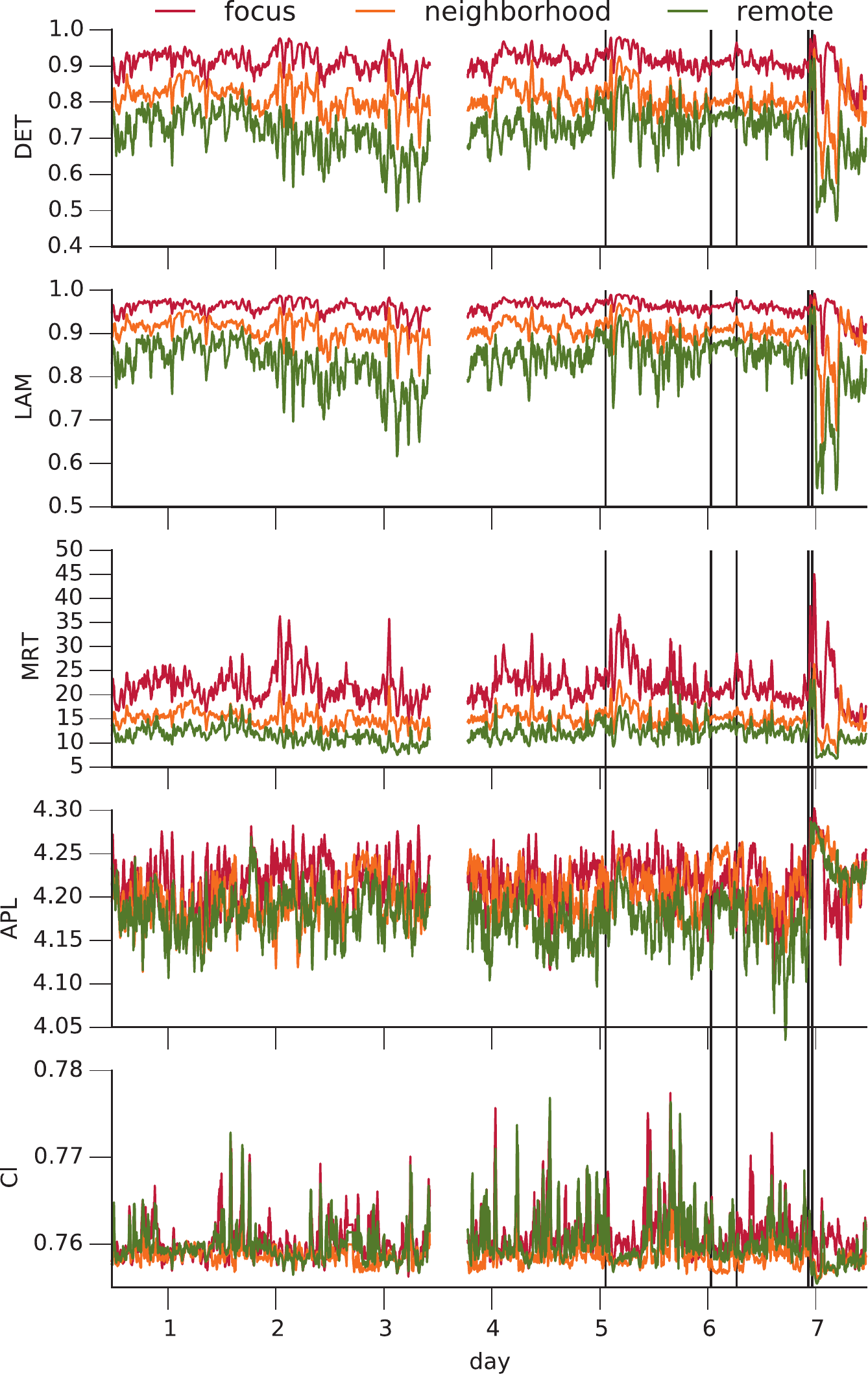}
\caption{\label{timeprodistributions_illus} 
Time profiles of the measures of complexity
DET, LAM, MRT, APL, and Cl (from top to bottom) estimated from iEEG data of patient 3. 
Data recorded from within the epileptic focus (red), from its neighborhood (orange), and from a remote recording site (green). 
Moving average over 60 windows corresponding to 20.48 min.
There was a recording gap on day 3.
Seizures are marked by black vertical lines, and tics on x-axes denote midnight.}
\end{figure}

In Fig.~\ref{timeprodistributions_illus} we show, as an example, time profiles of the MOC calculated from iEEG data of patient 3, which were recorded from within the epileptic focus (as defined by the pre-surgical workup), from its neighborhood (not more than two electrode contacts apart), and from a remote recording site. 
The temporal variability of the RQA-based measures DET and LAM appears to increase with an increasing distance to the epileptic focus, while the reverse appears to hold for MRT. 
The temporal means of these MOC decrease with an increasing distance to the epileptic focus,
and similar observations could be made for all other patients. 
This finding is in line with previous studies that employed other measures of complexity~\cite{Lehnertz1995,Weber1998,Widman2000,Lehnertz2001,Andrzejak2001,Andrzejak2006,Prusseit2007,Andrzejak2012,Naro2014,Subramaniyam2015}.
In contrast, no such clear dependence of the temporal mean could be observed for the RN-based measures APL and Cl, although their evolutions exhibit some periodic temporal structure that appears to be related to daily rhythms.
Indeed, estimating their power spectral densities (Lomb--Scargle periodogram~\cite{Press1989}) revealed a strong component at about \unit[24]{h} (data not shown) particularly for recordings from remote sites and near the epileptic focus when using APL and for recordings from remote sites and from within the epileptic focus when using Cl.
Similar observations could be made for all other patients, and comparable dependences on daily rhythms had been reported for global and local statistical measures of functional epileptic brain networks~\cite{kuhnert2010,lehnertz2014,geier2015}.

\begin{figure} [hhh!]
\centering
\includegraphics[width=\columnwidth]{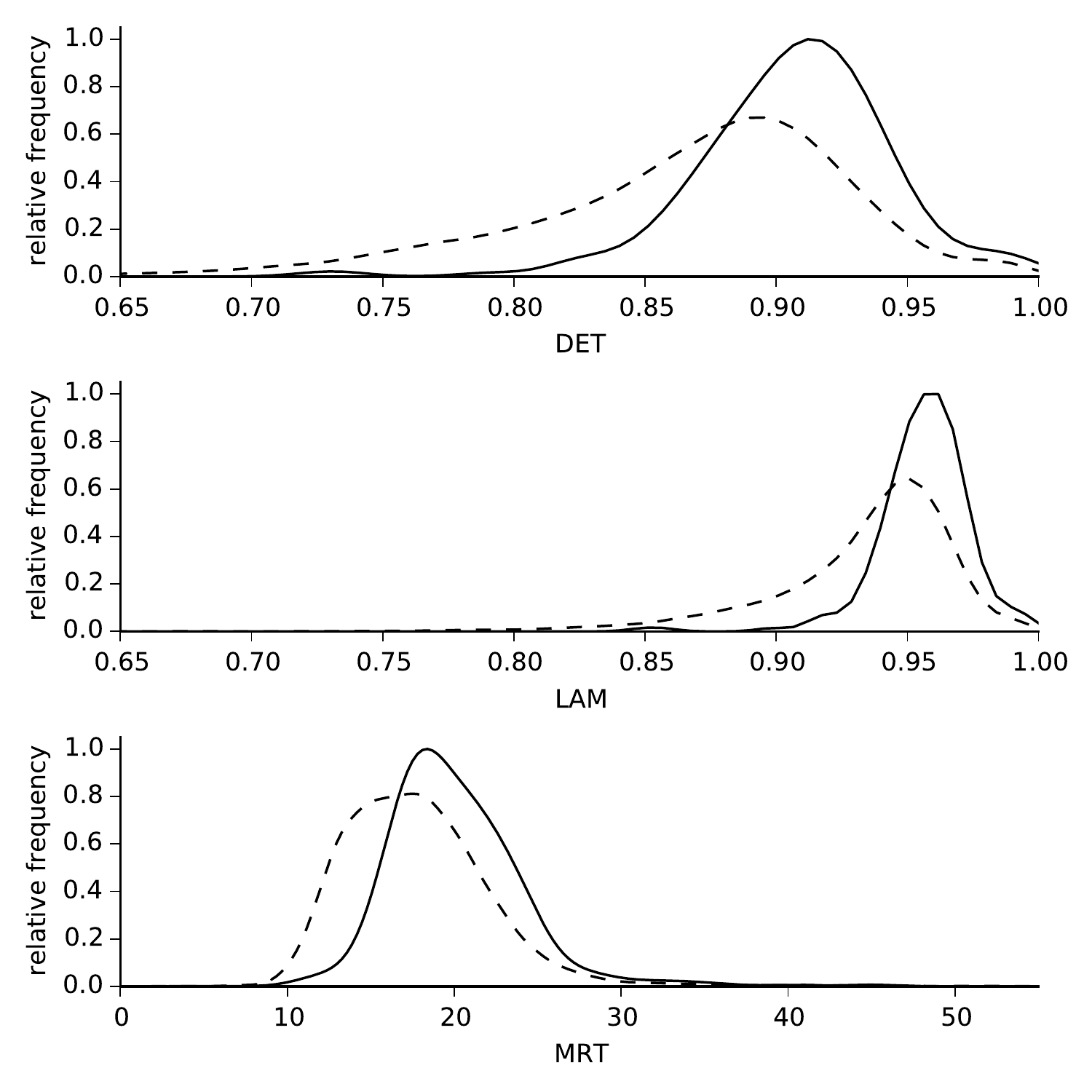}
\caption{\label{exemplarydistr} 
Frequency distributions of the inter-ictal (dashed lines) and pre-ictal (solid lines) values of the measures of complexity DET, LAM, and MRT (from top to bottom) estimated from iEEG data of patient 3. Data from a recording site near the epileptic focus carrying predictive information.}
\end{figure}

With the methods described in Sec.~\ref{sec:predmeth}, we observed statistically significant differences between MOC values from the pre-ictal and inter-ictal periods in three out of five patients.
In patient 1, both LAM and MRT indexed the dynamics of the same remote brain site (one recording site) to carry information predictive of impending seizures. 
Both MOC attained lower values during the \unit[30]{min} pre-ictal periods than during inter-ictal periods.
In patient 3, all RQA-based MOC provided predictive information (pre-ictally increased values, see Fig.\ref{exemplarydistr}) from the dynamics near the epileptic focus (two recording sites). 
In addition, APL identified other nearby brain sites (two recording sites) to carry potential seizure precursor dynamics (pre-ictally decreased values). 
With both APL and Cl remote but adjacent brain regions (one recording site each) could be identified, however, either through pre-ictally increased (APL) or decreased values (Cl).
In patient 4, all RQA-based MOC provided predictive information (pre-ictally decreased values) from the dynamics of the epileptic focus (one recording site), of nearby (one site) and of remote brain regions (five sites).
In addition, similar information could be achieved with Cl from the dynamics of remote brain regions (three sites). 

When judging the predictive performance of RQA- and RN-based measures, we note first that the RQA-based measures DET, LAM, and MRT pinpointed the same brain region to carry predictive information in all patients. 
Although these MOC assess different characteristics of an RP, they here provided redundant spatial information.
Nevertheless, seizure precursors were mostly found in brain regions off the epileptic focus, which would favor the recent concept of seizure generation in an {\it epileptic network} rather than from a circumscribed area of the brain (epileptic focus)~\cite{lehnertz2014}.
Pre-ictal alterations of these MOC (increase vs. decrease) exhibited a high interindividual variability, rendering an interpretation of the pre-seizure brain dynamics difficult. 

Second, RN-based measures APL and Cl identified precursory structures in two patients only, and particularly Cl clearly decreased during the \unit[30]{min} pre-ictal periods.
If Cl is interpreted as a global measure of the effective dimensionality of the underlying attractive set~\cite{donner2011}, its pre-ictal decline is in line with previous observations using other dimensionality estimates~\cite{Lehnertz1998}.
As with the RQA-based measures, APL and Cl also identified seizure precursors in brain regions off the epileptic focus.

Third, taking into account the redundancies provided by RQA-based MOC, both RQA- and RN-based measures had a comparable predictive performance, which compares to the one seen previously for other linear and nonlinear univariate measures~\cite{mormann2005}.

\section{Conclusion}

We investigated the suitability of selected measures of complexity based on recurrence quantification analysis (RQA) and recurrence networks (RN)---characterizing dynamical and geometrical properties of a system---for an identification of pre-seizure states in multi-day, multi-channel, invasive EEG recordings from five epilepsy patients.
We employed a number of downstream statistical techniques to avoid spurious findings due to various influencing factors and due to multiple comparisons.
With these approaches, statistically significant precursory structures could be identified in three patients. 
Findings indicate a high redundancy in predictive information that can be achieved with RQA-based measures. 
In the two patients for which RN-based measures identified precursory structures, these measures did not provide additional information about brain regions from which possible precursors emerge. 
We thus conclude that the combined use of both quantification approaches does not appear to improve the detection of dynamical changes preceding seizures in the human epileptic brain.
Clearly, a final judgment of the suitability of these recurrence-based time series analysis approaches for seizure prediction studies needs to be evaluated on a larger data base.

\section*{Acknowledgement}
We are grateful to {M.~Riedl and} G.~Ansmann for fruitful discussions and critical comments on earlier versions of the manuscript. This 
work was supported by the Volkswagen Foundation (Grant Nos. 88461, 88462, 88463, 85390, 85391 and 85392).

\appendix
\section{Measures of complexity}

The following measures of complexity have been used and their description is detailed in Refs.~\cite{marwan2007recurrence,ngamga2007,ngamga2012,marwan2009,donner2010}:

\begin{enumerate}

{\item {\it Determinism} (DET)} is the percentage of recurrence points which form a diagonal line of minimal length $l_{\rm min}$. 
It quantifies the predictability of the system. Processes with stochastic behavior will render a DET value which tends to 0, while it will be equal to 1 for purely periodic processes.
\begin{equation}\label{det}
\textrm{DET}=\frac{\sum_{l=l_{\rm min}}^T lP(l)}{\sum_{i,j=1}^T R_{i,j}},
\end{equation}
where $P(l)$ denotes the frequency distribution of the lengths $l$ of the diagonal lines in the RP.

{\item {\it Laminarity} (LAM) } is the percentage of recurrence points which form a black vertical line of minimal length $v_{\rm min}$. 
It represents slowly changing states and, thus, the occurrence of laminar states in the system. High values of LAM are an indication 
of dynamics that is trapped more often to certain states. 
\begin{equation}\label{lam}
\textrm{LAM}=\frac{\sum_{v=v_{\rm min}}^T vP(v)}{\sum_{v=1}^T vP(v)},
\end{equation}
where $P(v)$ denotes the frequency distributions of the lengths $v$ of black vertical lines.

{\item {\it Mean recurrence time} (MRT) } is the average length of white vertical lines in the RP.
\begin{equation}\label{mrt}
\textrm{MRT}=\frac{\sum_{w=1}^T wP(w)}{\sum_{w=1}^T P(w)},
\end{equation}
where $P(w)$ denotes the frequency distribution of the lengths $w$ of white vertical lines. MRT estimates the main time-scale of variations (e.g., for an harmonic oscillation it corresponds to the period length)~\cite{gao99}. 

In accordance with previous works~\cite{marwan2007recurrence}, the minimal lengths were chosen as $l_{\rm min}=2$ and $v_{\rm min}=2$.\\

As mentioned in section II, by considering the phase space vectors as nodes of a network and their recurrences in the phase space as links, an analogy between the recurrence matrix and the adjacency matrix of an undirected and unweighted network can be built. 
The network is represented by an adjacency matrix $\mathbf{A}$, which is the recurrence matrix from which the identity matrix is 
subtracted (${A}_{ij}={R}_{ij}-{\delta}_{ij}$; where ${\delta}_{ij}$ is the Kronecker delta used to avoid self-loops in the network). 

{\item {\it Average shortest path length} (APL)} is the average length of shortest paths between all pairs of nodes in the network and is given by:
\begin{equation}\label{apl}
\textrm{APL}=\frac{1}{T(T-1)}\sum_{i,j=1}^{T} d_{ij},
\end{equation}
with the minimum number of links $d_{ij}$ that have to be crossed to move from node $i$ to node $j$. 
Disconnected pairs of nodes are not included in the average. 
In recurrence networks, APL characterizes the average phase space separation of states~\cite{donner2010}.

{\item {\it Network transitivity} (Cl) } considers the average probability that two neighbors of any state are also neighbors and is given by: 
\begin{equation}\label{cc}
\textrm{Cl} = \frac{\sum_{i,j,k=1}^{T} A_{jk}A_{ij}A_{ik}}{\sum_{i,j,k=1}^{T} A_{ij}A_{ik}}.
\end{equation}

Cl can be interpreted as a global measure of the underlying attractive set's effective dimensionality~\cite{donner2011}.

\end{enumerate}


\begin{thebibliography}{10}
\expandafter\ifx\csname url\endcsname\relax
  \def\url#1{\texttt{#1}}\fi
\expandafter\ifx\csname urlprefix\endcsname\relax\def\urlprefix{URL }\fi
\expandafter\ifx\csname href\endcsname\relax
  \def\href#1#2{#2} \def\path#1{#1}\fi

\bibitem{eckmann87}
J.-P. Eckmann, S.~{Oliffson Kamphorst}, D.~Ruelle, Recurrence plots of
  dynamical systems, Europhysics Letters 4~(9) (1987) 973--977.
\newblock \href {http://dx.doi.org/10.1209/0295-5075/4/9/004}
  {\path{doi:10.1209/0295-5075/4/9/004}}.

\bibitem{marwan2007recurrence}
N.~Marwan, M.~C. Romano, M.~Thiel, J.~Kurths, Recurrence plots for the analysis
  of complex systems, Physics Reports 438~(5) (2007) 237--329.
\newblock \href {http://dx.doi.org/10.1016/j.physrep.2006.11.001}
  {\path{doi:10.1016/j.physrep.2006.11.001}}.

\bibitem{marwan2008epjst}
N.~Marwan, A historical review of recurrence plots, The European Physical
  Journal -- Special Topics 164~(1) (2008) 3--12.
\newblock \href {http://dx.doi.org/10.1140/epjst/e2008-00829-1}
  {\path{doi:10.1140/epjst/e2008-00829-1}}.

\bibitem{marwan2015a}
N.~Marwan, J.~Kurths, S.~Foerster, Analysing spatially extended
  high-dimensional dynamics by recurrence plots, Physics Letters A 379 (2015)
  894--900.
\newblock \href {http://dx.doi.org/10.1016/j.physleta.2015.01.013}
  {\path{doi:10.1016/j.physleta.2015.01.013}}.

\bibitem{zbilut1992}
J.~P. Zbilut, C.~L. {Webber, Jr.}, Embeddings and delays as derived from
  quantification of recurrence plots, Physics Letters A 171~(3--4) (1992)
  199--203.
\newblock \href {http://dx.doi.org/10.1016/0375-9601(92)90426-M}
  {\path{doi:10.1016/0375-9601(92)90426-M}}.

\bibitem{marwan2002heartrate}
N.~Marwan, N.~Wessel, U.~Meyerfeldt, A.~Schirdewan, J.~Kurths,
  Recurrence-plot-based measures of complexity and their application to
  heart-rate-variability data, Physical Review E 66~(2) (2002) 026702.
\newblock \href {http://dx.doi.org/10.1103/PhysRevE.66.026702}
  {\path{doi:10.1103/PhysRevE.66.026702}}.

\bibitem{webber2015}
C.~L. {Webber, Jr.}, N.~Marwan, Recurrence Quantification Analysis -- Theory
  and Best Practices, Springer, Cham, 2015.
\newblock \href {http://dx.doi.org/10.1007/978-3-319-07155-8}
  {\path{doi:10.1007/978-3-319-07155-8}}.

\bibitem{acharya2011}
U.~R. Acharya, S.~{Vinitha Sree}, S.~Chattopadhyay, W.~Yu, P.~C.~A. Ang,
  Application of recurrence quantification analysis for the automated
  identification of epileptic {EEG} signals, International Journal of Neural
  Systems 21~(3) (2011) 199--211.
\newblock \href {http://dx.doi.org/10.1142/S0129065711002808}
  {\path{doi:10.1142/S0129065711002808}}.

\bibitem{thomasson2001}
N.~Thomasson, T.~J. Hoeppner, C.~L. {Webber, Jr.}, J.~P. Zbilut, Recurrence
  quantification in epileptic EEGs, Physics Letters A 279~(1--2) (2001)
  94--101.
\newblock \href {http://dx.doi.org/10.1016/S0375-9601(00)00815-X}
  {\path{doi:10.1016/S0375-9601(00)00815-X}}.

\bibitem{komala2009}
C.~Komalapriya, M.~C. Romano, M.~Thiel, U.~Schwarz, J.~Kurths, J.~Simonotto,
  M.~Furman, W.~L. Ditto, P.~R. Carney, Analysis of high-resolution
  microelectrode {EEG} recordings in an animal model of spontaneous limbic
  seizures, International Journal of Bifurcation and Chaos 19~(2) (2009)
  605--617.
\newblock \href {http://dx.doi.org/10.1142/S0218127409023226}
  {\path{doi:10.1142/S0218127409023226}}.

\bibitem{zhang2008}
W.~Zhang, G.~Worrell, B.~He, Recurrence based deterministic trends in {EEG}
  records of epilepsy patients, in: Information Technology and Applications in
  Biomedicine, 2008. ITAB 2008. International Conference on, 2008, pp.
  391--394.
\newblock \href {http://dx.doi.org/10.1109/ITAB.2008.4570657}
  {\path{doi:10.1109/ITAB.2008.4570657}}.

\bibitem{zhu2008}
T.~Zhu, L.~Huang, S.~Zhang, Y.~Huang, Predicting epileptic seizure by
  recurrence quantification analysis of single-channel {EEG}, Lecture Notes in
  Computer Science: Advanced Intelligent Computing Theories and Applications.
  With Aspects of Theoretical and Methodological Issues 5226 (2008) 438--445.
\newblock \href {http://dx.doi.org/10.1007/978-3-540-87442-3}
  {\path{doi:10.1007/978-3-540-87442-3}}.

\bibitem{Thurman2011}
D.~J. Thurman, E.~Beghi, C.~E. Begley, A.~T. Berg, J.~R. Buchhalter, D.~Ding,
  D.~C. Hesdorffer, W.~A. Hauser, L.~Kazis, R.~Kobau, B.~Kroner, D.~Labiner,
  K.~Liow, G.~Logroscino, M.~T. Medina, C.~R. Newton, K.~Parko, A.~Paschal,
  P.-M. Preux, J.~W. Sander, A.~Selassie, W.~Theodore, T.~Tomson, S.~Wiebe, for
  the ILAE Commission~on Epidemiology, Standards for epidemiologic studies and
  surveillance of epilepsy, Epilepsia 52 (2011) 2--26.
\newblock \href {http://dx.doi.org/10.1111/j.1528-1167.2011.03121.x}
  {\path{doi:10.1111/j.1528-1167.2011.03121.x}}.

\bibitem{Litt2002b}
B.~Litt, K.~Lehnertz, Seizure prediction and the preseizure period, Current
  Opinion in Neurology 15 (2002) 173--177.

\bibitem{mormann2007}
F.~Mormann, R.~G. Andrzejak, C.~E. Elger, K.~Lehnertz, Seizure prediction: the
  long and winding road, Brain 130 (2007) 314--333.
\newblock \href {http://dx.doi.org/10.1093/brain/awl241}
  {\path{doi:10.1093/brain/awl241}}.

\bibitem{sackellares2008}
J.~C. Sackellares, Seizure prediction, Epilepsy Currents 8 (2008) 55--59.
\newblock \href {http://dx.doi.org/10.1111/j.1535-7511.2008.00236.x}
  {\path{doi:10.1111/j.1535-7511.2008.00236.x}}.

\bibitem{carney2011}
P.~R. Carney, S.~Myers, J.~D. Geyer, Seizure prediction: methods, Epilepsy
  Behavior 22 (2011) 94--101.
\newblock \href {http://dx.doi.org/doi: 10.1016/j.yebeh.2011.09.001}
  {\path{doi:doi: 10.1016/j.yebeh.2011.09.001}}.

\bibitem{tetzlaff2013}
R.~Tetzlaff, C.~E. Elger, K.~Lehnertz, Recent Advances in Predicting and
  Preventing Epileptic Seizures, World Scientific, 2013.

\bibitem{lehnertz2014}
K.~Lehnertz, G.~Ansmann, S.~Bialonski, H.~Dickten, C.~Geier, S.~Porz, Evolving
  networks in the human epileptic brain, Physica D: Nonlinear Phenomena 267
  (2014) 7--15.
\newblock \href {http://dx.doi.org/10.1016/j.physd.2013.06.009}
  {\path{doi:10.1016/j.physd.2013.06.009}}.

\bibitem{Gadhoumi2015}
K.~Gadhoumi, J.-M. Lina, F.~Mormann, J.~Gotman, Seizure prediction for
  therapeutic devices: A review, Journal of Neuroscience Methods (2015) --\href
  {http://dx.doi.org/10.1016/j.jneumeth.2015.06.010}
  {\path{doi:10.1016/j.jneumeth.2015.06.010}}.

\bibitem{Schuele2008}
S.~U. Schuele, H.~O. L\"uders, Intractable epilepsy: management and therapeutic
  alternatives, Lancet Neurology 7 (2008) 514--524.
\newblock \href {http://dx.doi.org/10.1016/S1474-4422(08)70108-X}
  {\path{doi:10.1016/S1474-4422(08)70108-X}}.

\bibitem{xu2008}
X.~Xu, J.~Zhang, M.~Small, Superfamily phenomena and motifs of networks induced
  from time series, Proceedings of the National Academy of Sciences 105~(50)
  (2008) 19601--19605.
\newblock \href {http://dx.doi.org/10.1073/pnas.0806082105}
  {\path{doi:10.1073/pnas.0806082105}}.

\bibitem{marwan2009}
N.~Marwan, J.~F. Donges, Y.~Zou, R.~V. Donner, J.~Kurths, Complex network
  approach for recurrence analysis of time series, Physics Letters A 373~(46)
  (2009) 4246--4254.
\newblock \href {http://dx.doi.org/10.1016/j.physleta.2009.09.042}
  {\path{doi:10.1016/j.physleta.2009.09.042}}.

\bibitem{donner2010}
R.~V. Donner, Y.~Zou, J.~F. Donges, N.~Marwan, J.~Kurths, {Recurrence networks
  -- a novel paradigm for nonlinear time series analysis}, New Journal of
  Physics 12~(3) (2010) 033025.
\newblock \href {http://dx.doi.org/10.1088/1367-2630/12/3/033025}
  {\path{doi:10.1088/1367-2630/12/3/033025}}.

\bibitem{marwan2015b}
N.~Marwan, J.~Kurths, Complex network based techniques to identify extreme
  events and (sudden) transitions in spatio-temporal systems, Chaos 25~(9)
  (2015) 097609.
\newblock \href {http://dx.doi.org/DOI:10.1063/1.4916924}
  {\path{doi:DOI:10.1063/1.4916924}}.

\bibitem{donner2015}
R.~V. Donner, J.~F. Donges, Y.~Zou, J.~H. Feldhoff, {Complex Network Analysis
  of Recurrences}, In: Recurrence Quantification Analysis -- Theory and Best Practices, Eds.:
  C.~L.~{Webber, Jr.} and N.~Marwan,
  Springer, Cham, 2015, pp. 101--163.
\newblock \href {http://dx.doi.org/10.1007/978-3-319-07155-8\_4}
  {\path{doi:10.1007/978-3-319-07155-8\_4}}.

\bibitem{zou2010}
Y.~Zou, R.~V. Donner, J.~F. Donges, N.~Marwan, J.~Kurths, Identifying complex
  periodic windows in continuous-time dynamical systems using recurrence-based
  methods, Chaos 20~(4) (2010) 043130.
\newblock \href {http://dx.doi.org/10.1063/1.3523304}
  {\path{doi:10.1063/1.3523304}}.

\bibitem{gao2013}
Z.~Gao, X.~Zhang, N.~Jin, N.~Marwan, J.~Kurths, Multivariate recurrence network
  analysis for characterizing horizontal oil-water two-phase flow, Physical
  Review E 88 (2013) 032910.
\newblock \href {http://dx.doi.org/10.1103/PhysRevE.88.032910}
  {\path{doi:10.1103/PhysRevE.88.032910}}.

\bibitem{feldhoff2012}
J.~H. Feldhoff, R.~V. Donner, J.~F. Donges, N.~Marwan, J.~Kurths, Geometric
  detection of coupling directions by means of inter-system recurrence
  networks, Physics Letters A 376~(46) (2012) 3504--3513.
\newblock \href {http://dx.doi.org/10.1016/j.physleta.2012.10.008}
  {\path{doi:10.1016/j.physleta.2012.10.008}}.

\bibitem{lang2013}
P.~Lang, D.-B. Liu, S.-M. Cai, L.~Hong, P.-L. Zhou, Recurrence network analysis
  of the synchronous EEG time series in normal and epileptic brains, Cell
  Biochemistry and Biophysics 66 (2013) 331--336.
\newblock \href {http://dx.doi.org/10.1007/s12013-012-9452-0}
  {\path{doi:10.1007/s12013-012-9452-0}}.

\bibitem{Horstmann2010}
M.-T. Horstmann, S.~Bialonski, N.~Noennig, H.~Mai, J.~Prusseit, J.~Wellmer,
  H.~Hinrichs, K.~Lehnertz, State dependent properties of epileptic brain
  networks: Comparative graph-theoretical analyses of simultaneously recorded
  {EEG} and {MEG}, Clinical Neurophysiology 121 (2010) 172--185.
\newblock \href {http://dx.doi.org/10.1016/j.clinph.2009.10.013}
  {\path{doi:10.1016/j.clinph.2009.10.013}}.

\bibitem{subramaniyam2014}
N.~P. Subramaniyam, J.~Hyttinen, Characterization of dynamical systems under
  noise using recurrence networks: {A}pplication to simulated and {EEG} data,
  Physics Letters A 378 (2014) 3464--3474.
\newblock \href {http://dx.doi.org/10.1016/j.physleta.2014.10.005}
  {\path{doi:10.1016/j.physleta.2014.10.005}}.

\bibitem{ramirez2013}
G.~M. {Ram\'{i}rez \'{A}vila}, A.~Gapelyuk, N.~Marwan, H.~Stepan, J.~Kurths,
  T.~Walther, N.~Wessel, Classifying healthy women and preeclamptic patients
  from cardiovascular data using recurrence and complex network methods,
  Autonomic Neuroscience 178~(1--2) (2013) 103--110.
\newblock \href {http://dx.doi.org/10.1016/j.autneu.2013.05.003}
  {\path{doi:10.1016/j.autneu.2013.05.003}}.

\bibitem{kral2002}
T.~Kral, H.~Clusmann, J.~Urbach, J.~Schramm, C.~E. Elger, M.~Kurthen,
  T.~Grunwald, Preoperative evaluation for epilepsy surgery ({Bonn Algorithm}),
  Zentralblatt f\"{u}r Neurochirurgie 63~(3) (2002) 106--110.
\newblock \href {http://dx.doi.org/10.1055/s-2002-35826}
  {\path{doi:10.1055/s-2002-35826}}.

\bibitem{goswami2013epjst}
B.~Goswami, N.~Marwan, G.~Feulner, J.~Kurths, {How do global temperature
  drivers influence each other? -- A network perspective using recurrences},
  The European Physical Journal -- Special Topics 222 (2013) 861--873.
\newblock \href {http://dx.doi.org/10.1140/epjst/e2013-01889-8}
  {\path{doi:10.1140/epjst/e2013-01889-8}}.

\bibitem{thiel2004a}
M.~Thiel, M.~C. Romano, P.~L. Read, J.~Kurths, {Estimation of dynamical
  invariants without embedding by recurrence plots}, Chaos 14~(2) (2004)
  234--243.
\newblock \href {http://dx.doi.org/10.1063/1.1667633}
  {\path{doi:10.1063/1.1667633}}.

\bibitem{Rieke2002}
C.~Rieke, K.~Sternickel, R.~G. Andrzejak, C.~E. Elger, P.~David, K.~Lehnertz,
  {Measuring Nonstationarity by Analyzing the Loss of Recurrence in Dynamical
  Systems}, Physical Review Letters 88 (2002) 244102.
\newblock \href {http://dx.doi.org/10.1103/PhysRevLett.88.244102}
  {\path{doi:10.1103/PhysRevLett.88.244102}}.

\bibitem{Rieke2003}
C.~Rieke, F.~Mormann, R.~G. Andrzejak, T.~Kreuz, P.~David, C.~E. Elger,
  K.~Lehnertz, Discerning nonstationarity from nonlinearity in seizure-free and
  preseizure {EEG} recordings from epilepsy patients, IEEE Trans. Biomed. Eng.
  50 (2003) 634--639.

\bibitem{Rieke2004}
C.~Rieke, R.~G. Andrzejak, F.~Mormann, K.~Lehnertz, Improved statistical test
  for nonstationarity using recurrence time statistics, Physical Review E 69
  (2004) 046111.
\newblock \href {http://dx.doi.org/10.1103/PhysRevE.69.046111}
  {\path{doi:10.1103/PhysRevE.69.046111}}.

\bibitem{marwan2011}
N.~Marwan, {How to avoid potential pitfalls in recurrence plot based data
  analysis}, {International Journal of Bifurcation and Chaos} 21~(4) (2011)
  1003--1017.
\newblock \href {http://dx.doi.org/10.1142/S0218127411029008}
  {\path{doi:10.1142/S0218127411029008}}.

\bibitem{ngamga2007}
E.~J. Ngamga, A.~Nandi, R.~Ramaswamy, M.~C. Romano, M.~Thiel, J.~Kurths,
  Recurrence analysis of strange nonchaotic dynamics, Physical Review E 75~(3)
  (2007) 036222.
\newblock \href {http://dx.doi.org/10.1103/PhysRevE.75.036222}
  {\path{doi:10.1103/PhysRevE.75.036222}}.

\bibitem{ngamga2012}
E.~J. Ngamga, D.~V. Senthilkurmar, A.~Prasad, P.~Parmananda, N.~Marwan,
  J.~Kurths, Distinguishing dynamics using recurrence-time statistics, Physical
  Review E 85~(2) (2012) 026217.
\newblock \href {http://dx.doi.org/10.1103/PhysRevE.85.026217}
  {\path{doi:10.1103/PhysRevE.85.026217}}.

\bibitem{donges2012}
J.~F. Donges, J.~Heitzig, R.~V. Donner, J.~Kurths, Analytical framework for
  recurrence network analysis of time series, Physical Review E 85 (2012)
  046105.
\newblock \href {http://dx.doi.org/10.1103/PhysRevE.85.046105}
  {\path{doi:10.1103/PhysRevE.85.046105}}.

\bibitem{mormann2005}
F.~Mormann, T.~Kreuz, C.~Rieke, R.~G. Andrzejak, A.~Kraskov, P.~David, C.~E.
  Elger, K.~Lehnertz, On the predictability of epileptic seizures, Clinical
  Neurophysiology 116 (2005) 569--587.
\newblock \href {http://dx.doi.org/10.1016/j.clinph.2004.08.025}
  {\path{doi:10.1016/j.clinph.2004.08.025}}.

\bibitem{kuhnert2010}
M.~T. Kuhnert, C.~E. Elger, K.~Lehnertz, Long-term variability of global
  statistical properties of epileptic brain networks, Chaos 20~(4) (2010)
  043126.
\newblock \href {http://dx.doi.org/10.1063/1.3504998}
  {\path{doi:10.1063/1.3504998}}.

\bibitem{blanco1995}
S.~Blanco, H.~Garcia, R.~{Quian Quiroga}, L.~Romanelli, O.~A. Rosso,
  Stationarity of the {EEG} series, Engineering in Medicine and Biology
  Magazine, IEEE 14 (1995) 395--399.
\newblock \href {http://dx.doi.org/10.1109/51.395321}
  {\path{doi:10.1109/51.395321}}.

\bibitem{silva1987}
F.~H. {Lopes da Silva}, {EEG} {A}nalysis: {T}heory and {P}ractice. in: {E.
  Niedermeyer}, {F. H. Lopes da Silva}, editors. {E}lectroencephalography,
  {B}asic principles, {C}linical applications and related fields, Baltimore,
  MD: Urban and Schwarzenberg (1987) 871--897.

\bibitem{schinkel2008}
S.~Schinkel, O.~Dimigen, N.~Marwan, {Selection of recurrence threshold for
  signal detection}, {The European Physical Journal -- Special Topics} 164~(1)
  (2008) 45--53.
\newblock \href {http://dx.doi.org/10.1140/epjst/e2008-00833-5}
  {\path{doi:10.1140/epjst/e2008-00833-5}}.

\bibitem{elger1998}
C.~E. Elger, K.~Lehnertz, Seizure prediction by non-linear time series analysis
  of brain electrical activity, European Journal of Neuroscience 10 (1998)
  786--789.
\newblock \href {http://dx.doi.org/10.1046/j.1460-9568.1998.00090.x}
  {\path{doi:10.1046/j.1460-9568.1998.00090.x}}.

\bibitem{levanquyen1999}
M.~{Le Van Quyen}, J.~Martinerie, M.~Baulac, F.~Valera, Anticipating epileptic
  seizures in real time by a non-linear analysis of similarity between {EEG}
  recordings, Neuroreport 10~(10) (1999) 2149--2155.
\newblock \href {http://dx.doi.org/10.1097/00001756-199907130-00028}
  {\path{doi:10.1097/00001756-199907130-00028}}.

\bibitem{andrzejak2003}
R.~G. Andrzejak, F.~Mormann, T.~Kreuz, C.~Rieke, A.~Kraskov, C.~E. Elger,
  K.~Lehnertz, Testing the null hypothesis of the nonexistence of a preseizure
  state, Physical Review E 67 (2003) 010901.
\newblock \href {http://dx.doi.org/10.1103/PhysRevE.67.010901}
  {\path{doi:10.1103/PhysRevE.67.010901}}.

\bibitem{benjamini1995}
Y.~Benjamini, Y.~Hochberg, Controlling the false discovery rate: a practical
  and powerful approach to multiple testing, Journal of the Royal Statistical
  Society: Series B 57~(1) (1995) 289--300.

\bibitem{Lehnertz1995}
K.~Lehnertz, C.~E. Elger, Spatio-temporal dynamics of the primary epileptogenic
  area in temporal lobe epilepsy characterized by neuronal complexity loss,
  Electroencephalography and Clinical Neurophysiology 95~(2) (1995) 108--117.

\bibitem{Weber1998}
B.~Weber, K.~Lehnertz, C.~E. Elger, H.~G. Wieser, Neuronal complexity loss in
  intracranial {EEG} recorded with foramen ovale electrodes predicts side of
  primary epileptogenic area in temporal lobe epilepsy: {A} replication study,
  Epilepsia 39 (1998) 922--927.
\newblock \href {http://dx.doi.org/10.1111/j.1528-1157.1998.tb01441.x}
  {\path{doi:10.1111/j.1528-1157.1998.tb01441.x}}.

\bibitem{Widman2000}
G.~Widman, K.~Lehnertz, H.~Urbach, C.~E. Elger, Spatial distribution of
  neuronal complexity loss in neocortical lesional epilepsies, Epilepsia 41
  (2000) 811--817.

\bibitem{Lehnertz2001}
K.~Lehnertz, R.~G. Andrzejak, J.~Arnhold, T.~Kreuz, F.~Mormann, C.~Rieke,
  G.~Widman, C.~E. Elger, Nonlinear {EEG} analysis in epilepsy: Its possible
  use for interictal focus localization, seizure anticipation, and prevention,
  Journal of Clinical Neurophysiology 18~(3) (2001) 209--222.
\newblock \href {http://dx.doi.org/10.1097/00004691-200105000-00002}
  {\path{doi:10.1097/00004691-200105000-00002}}.

\bibitem{Andrzejak2001}
R.~G. Andrzejak, G.~Widman, K.~Lehnertz, P.~David, C.~E. Elger, The epileptic
  process as nonlinear deterministic dynamics in a stochastic environment: An
  evaluation on mesial temporal lobe epilepsy, Epilepsy Res. 44 (2001)
  129--140.
\newblock \href {http://dx.doi.org/10.1016/S0920-1211(01)00195-4}
  {\path{doi:10.1016/S0920-1211(01)00195-4}}.

\bibitem{Andrzejak2006}
R.~G. Andrzejak, F.~Mormann, G.~Widmann, T.~Kreuz, C.~E. Elger, K.~Lehnertz,
  Improved spatial characterization of the epileptic brain by focusing on
  nonlinearity, Epilepsy Research 69 (2006) 30--44.

\bibitem{Prusseit2007}
J.~Prusseit, K.~Lehnertz, Stochastic qualifiers of epileptic brain dynamics,
  Physical Review Letters 98 (2007) 138103.
\newblock \href
  {http://dx.doi.org/http://dx.doi.org/10.1103/PhysRevLett.98.138103}
  {\path{doi:http://dx.doi.org/10.1103/PhysRevLett.98.138103}}.

\bibitem{Andrzejak2012}
R.~G. Andrzejak, K.~Schindler, C.~Rummel, Nonrandomness, nonlinear dependence,
  and nonstationarity of electroencephalographic recordings from epilepsy
  patients, Physical Review E 86 (2012) 046206.
\newblock \href {http://dx.doi.org/10.1103/PhysRevE.86.046206}
  {\path{doi:10.1103/PhysRevE.86.046206}}.

\bibitem{Naro2014}
D.~Naro, C.~Rummel, K.~Schindler, R.~G. Andrzejak, Detecting determinism with
  improved sensitivity in time series: Rank-based nonlinear predictability
  score, Physical Review E 90 (2014) 032913.
\newblock \href {http://dx.doi.org/10.1103/PhysRevE.90.032913}
  {\path{doi:10.1103/PhysRevE.90.032913}}.

\bibitem{Subramaniyam2015}
N.~P. Subramaniyam, J.~Hyttinen, Dynamics of intracranial
  electroencephalographic recordings from epilepsy patients using univariate
  and bivariate recurrence networks, Physical Review E 91 (2015) 022927.
\newblock \href {http://dx.doi.org/10.1103/PhysRevE.91.022927}
  {\path{doi:10.1103/PhysRevE.91.022927}}.

\bibitem{Press1989}
W.~H. Press, G.~B. Rybicki, Fast algorithm for spectral analysis of unevenly
  sampled data, Astrophysical Journal 338 (1989) 277--280.
\newblock \href {http://dx.doi.org/10.1086/167197} {\path{doi:10.1086/167197}}.

\bibitem{geier2015}
C.~Geier, K.~Lehnertz, S.~Bialonski, Time-dependent degree-degree correlations
  in epileptic brain networks: from assortative to dissortative mixing,
  Frontiers in Human Neuroscience 9 (2015) 462.
\newblock \href {http://dx.doi.org/10.3389/fnhum.2015.00462}
  {\path{doi:10.3389/fnhum.2015.00462}}.

\bibitem{donner2011}
R.~V. Donner, J.~Heitzig, J.~F. Donges, Y.~Zou, N.~Marwan, J.~Kurths, The
  geometry of chaotic dynamics -- a complex network perspective, The European
  Physical Journal B 84 (2011) 653--672.
\newblock \href {http://dx.doi.org/10.1140/epjb/e2011-10899-1}
  {\path{doi:10.1140/epjb/e2011-10899-1}}.

\bibitem{Lehnertz1998}
K.~Lehnertz, C.~E. Elger, Can epileptic seizures be predicted? {Evidence} from
  nonlinear time series analysis of brain electrical activity, Physical Review
  Letters 80 (1998) 5019--5023.
\newblock \href {http://dx.doi.org/10.1103/PhysRevLett.80.5019}
  {\path{doi:10.1103/PhysRevLett.80.5019}}.

\bibitem{gao99}
J.~B. Gao, Recurrence time statistics for chaotic systems and their
  applications, Physical Review Letters 83~(16) (1999) 3178--3181.
\newblock \href {http://dx.doi.org/10.1103/PhysRevLett.83.3178}
  {\path{doi:10.1103/PhysRevLett.83.3178}}.

\end{thebibliography}

\end{document}